\documentclass[aps,prl,superscriptaddress,preprintnumbers,twocolumn,
 showpacs,twoside]{revtex4}

\usepackage{graphicx}% Include figure files
\usepackage{dcolumn}% Align table columns on decimal point
\usepackage{bm}% bold math
\usepackage{amsfonts}
\usepackage{amsmath}

\begin{document}

\title{Ising transition in the two-dimensional quantum $J_1-J_2$ 
Heisenberg model} 

\author{Luca Capriotti}
%\email{caprio@kitp.ucsb.edu}
\affiliation{Kavli Institute for Theoretical Physics, University of California,
Santa Barbara, CA 93106-4030}
\author{Tommaso Roscilde}
%\email{roscilde@usc.edu}
\affiliation{Department of Physics and Astronomy, University of Southern
California, Los Angeles, CA 90089-0484}
\author{Andrea Fubini}
%\email{fubini@fi.infn.it}
\affiliation{Dipartimento di Fisica, Universit\`a di Firenze, and INFM
UdR Firenze, Via G. Sansone 1, I-50019 Sesto Fiorentino (FI), Italy}
\author{Valerio Tognetti}
%\email{tognetti@fi.infn.it}
\affiliation{Dipartimento di Fisica, Universit\`a di Firenze, and INFM
UdR Firenze, Via G. Sansone 1, I-50019 Sesto Fiorentino (FI), Italy}
\date{\today}

\begin{abstract}
We study the thermodynamics of the spin-$S$
two-dimensional quantum Heisenberg antiferromagnet
on the square lattice with nearest ($J_1$) and next-nearest ($J_2$)
neighbor couplings in its collinear phase ($J_2/J_1>0.5$), 
using the pure-quantum self-consistent harmonic 
approximation. Our results show the persistence of a 
finite-temperature Ising phase transition 
%- marked by a divergence in the specific heat - 
for every value of the spin, 
provided that the ratio $J_2/J_1$ is greater than
a critical value corresponding to the onset of collinear long-range order 
at zero temperature.
We also calculate the spin- and temperature-dependence of
the collinear susceptibility and correlation length, and we discuss 
our results in light of the experiments on Li$_2$VOSiO$_4$ and related 
compounds.
\end{abstract}

% 75.10.Jm - Quantized spin models
% 05.30.-d - Quantum statistical mechanics
% 75.40.-s - Critical-point effects, specific heats, short range order
% 75.40.Cx - Static properties

\maketitle

The study of frustrated quantum spin systems is one of the 
most challenging and exciting topics in theoretical magnetism. 
A very extensively investigated, yet largely debated model is 
the so-called $J_1{-}J_2$ Heisenberg model with
competing antiferromagnetic couplings ($J_1,J_2>0$) 
between nearest-neighbors ($n.n$) and next-nearest-neighbors ($n.n.n.$)
\begin{equation}
 \hat{\cal H} = J_1 \sum_{n.n.}
 \hat{\bm S}_{\bm i}\cdot\hat{\bm S}_{\bm j}+ 
 J_2 \sum_{n.n.n.}
 \hat{\bm S}_{\bm i}\cdot\hat{\bm S}_{\bm j}~,
 \label{e.j1j2}
 \end{equation} 
\noindent
where $\hat{\bm S}_{\bm i}$ are spin-$S$ operators on a periodic
lattice with $N=L\times L$ sites; hereafter $\alpha = J_2/J_1$ 
defines the frustration ratio.

In the classical limit ($S\to\infty$), the minimum energy
configuration has conventional N\'eel order with magnetic wave vector
$\bm Q = (\pi,\pi)$ for $\alpha < 0.5$. Instead, for $\alpha > 0.5$,
the antiferromagnetic order is established independently on the two
sublattices, with the two (staggered) magnetizations free to rotate
with respect to each other. Among this degenerate manifold, two
families of {\it collinear states}, with pitch vectors $\bm Q =
(\pi,0)$ or $(0,\pi)$, are selected by an {\it order-by-disorder}
mechanism as soon as thermal or quantum fluctuations are taken into
account.  As a result, for $\alpha > 0.5$ the classical ground state
breaks not only the spin rotational and translational invariance of
the Hamiltonian -- as the conventional N\'eel phase -- but also its
invariance under $\pi/2$ lattice rotations, the resulting degeneracy
corresponding to the group $O(3)\times Z_2$. Remarkably, the
additional {\em discrete} $Z_2$ symmetry can in principle be broken at
finite temperatures without violating the Mermin-Wagner
theorem~\cite{mermin}, which applies in two dimensions only to
continuous ones. On this basis, in a seminal paper~\cite{ccl},
Chandra, Coleman and Larkin (CCL) proposed that the two-dimensional
$J_1-J_2$ model could sustain a Ising phase transition at
finite-temperature, with an order parameter directly related to the
$Z_2$ degree of freedom induced by frustration.  They also provided
quantitative estimates of the critical temperatures in the 
large-$\alpha$ limit for both the classical and the quantum cases.

While the CCL transition in the classical model has been recently
established by an extensive Monte Carlo (MC) study~\cite{weber}, the
occurrence of a low-temperature phase with a discrete broken symmetry
in the quantum case is still a subject of debate.  Besides its own
theoretical interest, this issue has become particularly important in
connection with the discovery of three vanadate compounds (${\rm
Li_2VOSiO_4}$, ${\rm Li_2VOGeO_4}$, and ${\rm VOMoO_4}$) whose
relevant magnetic interactions involve nearest and next-nearest
spin-$1/2$ $V^{4+}$ ions on weakly coupled stacked
planes~\cite{melzi,carretta}.  In particular, NMR and $\mu$SR
measurements on ${\rm Li_2VOSiO_4}$~\cite{melzi} indicate the
occurrence of a transition to a low-temperature phase with collinear
order at $T_{_{\rm N}}\simeq 2.8$ K. The selection of the collinear
ground state suggests the CCL mechanism as possible underlying
explanation.  Unfortunately, a clear experimental and theoretical
picture is still elusive: in the experiments with vanadate compounds,
structural distortions, interlayer and anisotropy effects are likely to
come into play~\cite{melzi}, and on the other hand the theoretical
investigation cannot rely on the insight provided by quantum Monte
Carlo methods as their reliability in presence of frustration is
strongly limited by the infamous {\em sign problem}~\cite{revgs}.

The necessary condition for the CCL transition to take place is the
presence of collinear order at zero temperature.  To this respect, the
existent theoretical results~\cite{revgs} point towards a
collinear-ground state for frustration ratios $\alpha>\alpha_c$, with
the critical value, $\alpha_c$, increasing as the value of the spin
decreases (for $S=1/2$, $\alpha_c \simeq 0.6$). Below this value
quantum fluctuations seem to be strong enough to stabilize a
low-temperature phase with short-range magnetic
correlations~\cite{pig}, but above it a CCL transition is in principle
possible.  However, this possibility has been recently challenged by
various high-temperature expansion
studies~\cite{rosener,misguich,singh} which were not able to detect
any evidence of a finite-temperature transition for $S=1/2$. On this
basis, Singh {\em et al.}~\cite{singh} have recently proposed a
scenario where, due to quantum fluctuations, the broken lattice
symmetry of the collinear ground state would be restored at any
non-zero temperature.

In this paper, we present a complete study of the thermodynamic
properties of the the quantum $J_1{-}J_2$ model in its collinear phase,
obtained within an effective Hamiltonian approach: the 
{\it pure-quantum self-consistent harmonic approximation} 
(PQSCHA)~\cite{Cuccolietal95}. This approach, based on to the 
path-integral formalism, allows one to separate the classical from the
pure-quantum contribution to the thermodynamics of the system.
Both the classical physics and the purely quantum linear 
effects are exactly described within the PQSCHA at {\em any} temperature, 
while the purely quantum non-linear contributions are
treated within a self-consistent harmonic approximation.
This feature makes the PQSCHA a valid tool to 
investigate the effects of quantum fluctuations on 
phase transitions -- like the CCL one -- whose character 
is essentially classical. In particular, this approach has been 
successfully applied to a variety of spin systems displaying 
Kosterlitz-Thouless and/or Ising critical behaviors, providing
reliable estimates of the transition temperatures
even for $S=1/2$~\cite{pqschacritical}.

Within the PQSCHA framework, the thermodynamics of a quantum
system is rephrased in terms of a classical effective Hamiltonian
with renormalized parameters depending on 
the spin value, temperature, and frustration. 
The derivation of the effective Hamiltonian for the $J_1{-}J_2$ model
closely follows the steps shown in Ref.~\cite{Cuccolietal95}.
The only detail which is worth mentioning here is that, in this case, 
the calculation of quantum renormalizations, involving a harmonic expansion
around one of the two families of collinear states, 
gives rise in general to solutions with an explicitly broken
symmetry under $\pi/2$ lattice rotations. 
However, it is possible to show that, to 
$O(1/S)$, the effective Hamiltonian can be recast 
in a form preserving all the symmetries of the original model,
and that reads (except for uniform terms):
\begin{equation}
{\cal H}^{\rm eff}=J_1^{\rm eff}\tilde{S}^2  \sum_{n.n.}  {\bm
s}_{\bm i}\cdot{\bm s}_{\bm j}~+~ J_2^{\rm eff} \tilde{S}^2  
\sum_{n.n.n.}  {\bm s}_{\bm i}\cdot {\bm s}_{\bm j} ~,
\label{e.j1j2eff}
\end{equation} 
where ${\bm s}_{\bm i}$ are classical vectors of length 1,  $\tilde S = S +
\frac12$ is the effective spin length~\cite{Cuccolietal95}, and
$J^{\rm eff}_1 = (\theta_x^2 + \theta_y^2)\theta_2^2 J_1/2$,
$J^{\rm  eff}_2 = \theta_2^4 J_2$,
are the quantum-renormalized exchange integrals, with spin-,
temperature- and frustration-dependent renormalization parameters
$\theta_x$, $\theta_y$ and $\theta_2$.
These, {\it e.g.} referred to the ground state with $\bm Q = (\pi,0)$,
are given by $\theta_\alpha^{2} = 1 - {{\cal D}_\alpha}/{2}$
($\alpha=x,y,2$) where the coefficients
\begin{equation}
{\cal D}_\alpha =  \frac{1}{N\tilde{S}} \sum_{\bm k} 
\frac{a^+_{\bm k}}{a^-_{\bm k}} 
\left(1- \gamma_{\bf k}^{(\alpha)} \right) {\cal L}_{\bm k} 
\end{equation}
are self-consistently determined with 
\begin{align}
\frac{a_{\bm k}^{\pm}}{2\tilde{S}\sqrt{J_1}}  &=  
\sqrt{\alpha\theta_2^2 
\left(1 \pm \gamma_{\bm k}^{(2)} \right)
+ \frac{\theta_{xy}^{2}}{2} \pm
 \theta_x^2 \gamma_{\bm k}^{(x)} +  
 \theta_y^2 \gamma_{\bm k}^{(y)}}~, \notag \\
{\cal L}_{\bm k} & = \coth f_{\bm k} - \frac{1}{f_{\bm k}}~,~~~~~
f_{\bm k} = \frac{\hbar \omega_{\bm k}}{2 \tilde{S} k_{\rm B} T}~,
\end{align}
where $\theta_{xy}^{2} = \theta_x^{2} - \theta_y^{2}$, $\gamma_{\bm
k}^{(x,y)} = \cos k_{x,y}/2$, $\gamma_{\bm k}^{(2)} = \cos k_x \cos
k_y$, $T$ is the temperature, and 
the renormalized dispersion relation is $\omega_{\bm k} =
a^+_{\bm k} a^-_{\bm k}$. 
Interestingly, at $T=0$ the PQSCHA turns out to be equivalent to the
modified spin-wave theory~\cite{mswt,singh} which is expected to be a
faithful representation of the low-energy properties in the collinear
phase.  In particular, the spin-wave dispersion relation $\omega_{\bm
k}$ has been recently shown to be in remarkable quantitative agreement
with series expansion results in the entire range of momenta for
$\alpha > \alpha_c$, and it is therefore expected to provide an
accurate description of the low-energy excitations of the
model~\cite{singh}.

\begin{figure}
\includegraphics[width=72mm,angle=0]{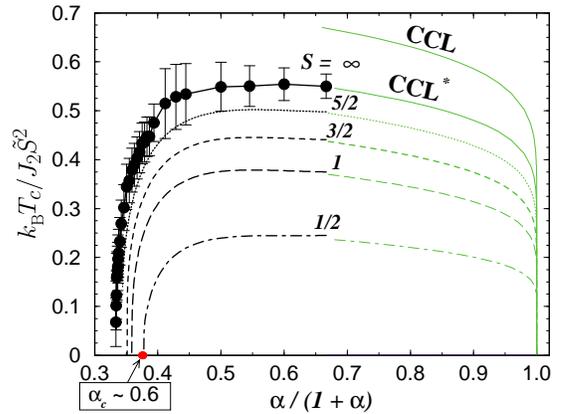}
 \caption{\label{f.tc} Renormalized critical temperature of the CCL
 transition for various values of the spin evaluated using Eq.(\ref{e.tccl})
 (non-solid lines). Classical data ($\bullet$) are taken from
 Ref.~\onlinecite{weber}. The solid lines on the right are the CCL and
 the CCL$^*$ prediction for the classical case (see text). The arrow
 marks the boundary of the non-magnetic phase for
 $S=1/2$~\cite{note}.}
\end{figure}  
 
The occurrence of the CCL transition in the quantum case can be
directly addressed within our approach by calculating the critical
temperatures as functions of the spin and of the frustration ratio: a
zero value of the critical temperature, or the breakdown of the
self-consistent harmonic treatment of quantum fluctuations, would
signal a possible absence of the phase transition. In particular,
using a simple scaling argument the critical temperatures in the
quantum case $T_c(S,\alpha)$ can be related to those of the classical
model $T^{(cl)}_c(\alpha)$ through the following self-consistent
relation~\cite{pqschacritical}
\begin{equation}
T_{c}(S,\alpha) = j_1^{\rm eff}(T_{c},S,\alpha)~
T^{(cl)}_c(\alpha^{\rm eff}(T_c,S,\alpha))~,
\label{e.tccl}
\end{equation}
where $j_1^{\rm eff} =  J_1^{\rm eff} \tilde S^2 /J_1$ and $\alpha^{\rm eff} =
J_2^{\rm eff}/J_1^{\rm eff}$. 
The classical transition temperature, $T^{(cl)}_c(\alpha)$ 
is accurately known through extensive MC simulations for $\alpha\leq 2$;
it vanishes for $\alpha\to 1/2$ and grows more or less linearly for $\alpha>1$.
Beyond $\alpha\simeq 2$ the determination of the critical temperatures 
becomes troublesome due to severe finite-size effects related to the
width of the domain walls between domains with
${\bm Q}=(\pi,0)$ and ${\bm Q}=(0,\pi)$~\cite{weber}. However,
for large values of the frustration ratio the classical CCL estimate
of the critical temperature, $T_c=0.768J_2/[1+0.135\ln(J_2/J_1)]$,  
is expected to be reliable~\cite{ccl,weber}. This vanishes 
logarithmically in the limit $J_2/J_1\to \infty$, corresponding to two 
decoupled unfrustrated Heisenberg systems. 
\begin{figure}
\includegraphics[width=70mm,angle=0]{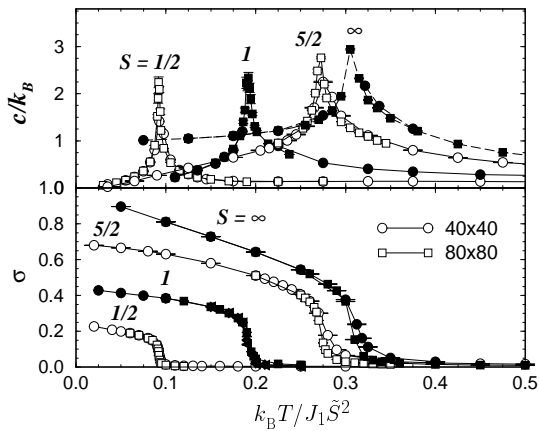}
 \caption{\label{f.trans} PQSCHA results for the specific heat (upper panel) 
 and the order parameter $\sigma$ (lower panel) as function of 
 temperature for various spins, in the case $\alpha = 0.65$.} 
\end{figure}  
In order to represent the whole interval of $\alpha\in [1/2,\infty)$
in Fig.~\ref{f.tc} we have plotted both the MC and the CCL estimates
of the classical critical temperatures as a function of
$\alpha/(1+\alpha)$. The mismatch between the MC and CCL predictions
is a minor flaw that can be easily accounted for, and corrected by
slightly modifying CCL's criterion for the determination of the
transition temperature as explained in Ref.~\cite{weber}. This gives
rise to the curve marked in Fig.~\ref{f.tc} as CCL$^*$. Using the
latter estimate, and the classical MC results of Ref.~\cite{weber},
the transition temperatures in the quantum case can be determined by
numerically solving Eq.~(\ref{e.tccl}). As shown in Fig.~\ref{f.tc},
the critical temperatures decrease as the spin decreases, due to the
enhancement of quantum fluctuations. Remarkably, while for large
$\alpha$ the transition temperature vanishes for $\alpha \to \infty$
for any value of the spin, in the opposite limit the critical
temperatures vanish approaching a critical value $\alpha_c>0.5$ that
increases as $S$ decreases, thus confirming the existence of a
non-magnetic phase in the regime of high frustration. 
In particular, for $S=1/2$, $\alpha_c\simeq 0.6$ in
agreement with the previous estimates of the zero-temperature quantum
critical point~\cite{revgs}.  For large $\alpha$, the PQSCHA results
turn out to be consistent with the CCL estimates for the quantum
$S=1/2$ system~\cite{ccl,weber}.  For intermediate values of $\alpha$,
the transition temperatures remain finite for any spin value. However
for $S=1/2$, though sizeable ($T_c/J_2\simeq0.2$), the critical
temperatures are more than an order of magnitude smaller than those
considered in Ref.~\cite{singh} in the discussion of the
high-temperature expansions results leading to the proposal of a $T=0$
critical scenario. On the contrary, a finite-temperature phase
transition at the critical temperatures we estimate is in fact
consistent with the numerical results of Ref.~\cite{singh}.

The study of the thermodynamics of the effective classical model
(\ref{e.j1j2eff}) provides a deeper analysis of the CCL transition. To
this end, we have performed classical MC simulations on the effective
Hamiltonian on $L\times L$ lattices, with $L$ up to 300. The PQSCHA
expression for the order parameter associated to the CCL
transition~\cite{ccl} is $\sigma = \theta_2^2~(\theta_x^2+\theta_y^2)
\langle | ( {\bm s}_{1}-{\bm s}_{3})\cdot ( {\bm s}_{2}-{\bm s}_{4}) |
\rangle_{\rm eff}/8$, where $(1,2,3,4)$ are the sites of on an
elementary plaquette of the lattice with counterclockwise numbering,
and $\langle...\rangle_{\rm eff}$ is the statistical average
associated to the effective classical Hamiltonian
(\ref{e.j1j2eff}). Another quantity bearing clear signatures of the
transition is the specific heat, calculated here as the numeric
derivative of the internal energy per spin, $c = \partial u/ \partial
T$, where $u=\langle{\cal H}_{\rm eff}\rangle/N - J_1 \tilde S^2 (
\theta_x^4 -\theta_y^4)$. These quantities are shown for $\alpha=0.65$
in Fig.~\ref{f.trans}. The Ising transition is clearly marked by a
(finite-size) peak in the specific heat in correspondence to the
temperature where the Ising order parameter vanishes. By decreasing
the spin, the saturated values of $\sigma$ decrease as well, due to
the increased quantum fluctuations, and so does the peak of the
specific heat, making the transition in the quantum case generally
weaker than in the classical one.  When $\alpha$ is increased, the
signatures of the transition are even more dramatically suppressed,
due to the mentioned effects related to the width of the domain
walls~\cite{weber}.
\begin{figure}
\includegraphics[width=65mm,angle=0]{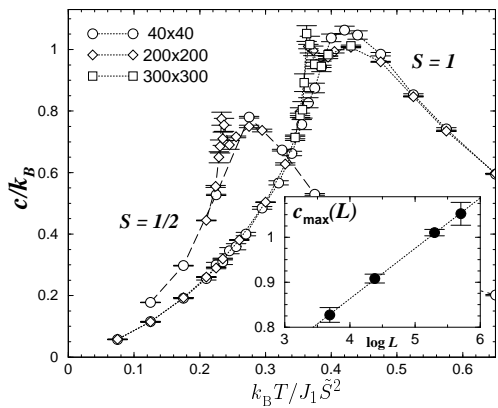}
 \caption{\label{f.cvS1alpha1} PQSCHA results for the specific heat 
 in the cases $\alpha=1$, $S=1,1/2$. In the inset: scaling of the 
 specific heat at the peak temperature $k_{\rm B}T/J_1\tilde{S}^2= 0.36$ 
 for $S=1$.} 
\end{figure}  
The Heisenberg features become more and more important with respect to
the frustration effects, in fact, as shown in Fig.~\ref{f.cvS1alpha1},
already for $\alpha = 1$, the Ising peak of the specific heat tends to
be masked by the broad maximum reminiscent of the unfrustrated limit
$\alpha \to \infty$, and very large values of $L$ are required in
order to resolve the logarithmic divergence (inset).  Similarly, the
absence of critical features in the recent numerical calculations of
the specific heat~\cite{misguich} can be traced back to the limited
range of the correlators generated by the high-temperature
expansion~\cite{privmisg}.

\begin{figure}
\includegraphics[width=68mm]{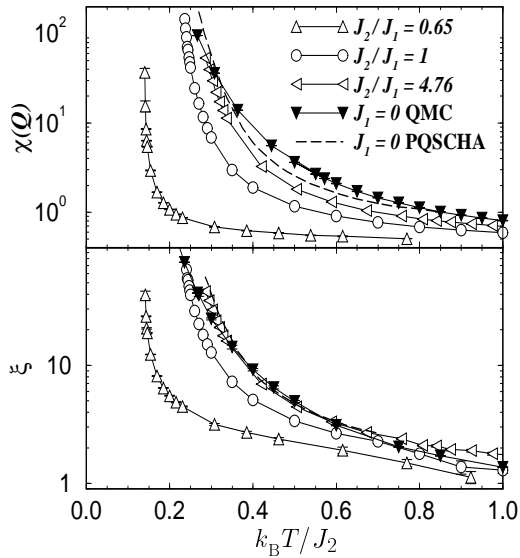}
 \caption{\label{f.chixi} Upper panel: $S=1/2$ staggered
 susceptibility, with ${\bm Q}=(\pi,0)$ and $(\pi,\pi)$ in the
frustrated (open symbols) and unfrustrated cases,
respectively. Lower
panel: $S=1/2$ correlation length. The data for the unfrustrated cases are
 taken from Ref.~\onlinecite{Cuccolietal95} and partly
 from Ref.~\onlinecite{KimT98}.} 
\end{figure}
  
In order to investigate the effect of frustration on the antiferromagnetic
correlations, we have also calculated the collinear susceptibility 
\begin{equation}
\chi({\bf Q})=\frac{S(S+1)}{3}+ \frac{\theta_0^4 \tilde{S}^2}{3}
 \sum_{\bm r\ne 0} e^{i\bm Q\cdot \bm r}
  \langle {\bm s}_{\bm 0}\cdot {\bm s}_{\bm r}\rangle_{\rm eff}~,
\end{equation}
with ${\bf Q}=(\pi,0)$ or $(0,\pi)$, and the corresponding
correlation length (through
the second moment estimator~\cite{Cooperetal82}). The 
results for the case $S=1/2$ for different values of $\alpha$
are shown in Fig. \ref{f.chixi}.
We observe that both quantities are strongly suppressed by
frustration for $\alpha \lesssim 1$, but for larger $\alpha$
the results are extremely close to those in the 
unfrustrated Heisenberg model, corresponding to $\alpha\to \infty$.
Therefore neutron scattering experiments on 
Li$_2$VOSiO$_4$ and related compounds with a large $\alpha$
are not likely to show any strong frustration effects 
in the magnetic correlations, as also 
indicated by preliminary experimental
results~\cite{neutron}.

Finally, we compare the estimated critical temperature for $S=1/2$
with the transition temperature, $T_{_{\rm N}} = 2.8$~K, observed in
Li$_2$VOSiO$_4$.  For this compound, various ratios $J_2/J_1$ have
been estimated, ranging from $J_2/J_1$~=~1.1 with $J_1 =
3.9$~K~\cite{melzi} to $J_2/J_1 =$~4.76 with $J_1 =
1.25$~K~\cite{misguich}.  Using these estimates, we get $T_c \approx
1.01$~K and 1.46~K, respectively, which are well below the transition
to three-dimensional collinear order observed at $T_{_{\rm N}}\simeq
2.8$~K. The CCL critical behavior is therefore not detectable 
in this compound~\cite{melzi}. 
In general, due to the weak nature of the
transition for $J_2/J_1\gtrsim 1$, previously discussed, the
observation of the CCL transition would require realizations of a
$J_1-J_2$ model with a smaller value of the ratio $J_2/J_1$.

In conclusion, using the pure-quantum self-consistent harmonic
approximation, we have provided a complete and consistent picture of
the collinear phase of the 2D quantum $J_1-J_2$ antiferromagnet.  Our
results indicate that the finite temperature transition predicted by
Chandra, Coleman, and Larkin~\cite{ccl} persists down to $S=1/2$
provided that the ratio $J_2/J_1$ is greater than a critical value
corresponding to the stabilization of collinear long-range order in
the ground state.  We believe that our findings will be a valid
reference point for future experimental investigations on
Li$_2$VOSiO$_4$ and related compounds.

We thank F.~Becca, P.~Carretta, A.~Cuccoli, G.~Misguich, and F.~Mila 
for fruitful discussion and correspondence.
This work was supported by NSF under Grant No. DMR02-11166 
(L.C.), DOE under Grant No. DE-FG03-01ER45908 (T.R.), 
and by INFN, INFM and MIUR-COFIN2002 (A.F. and V.T.).


\begin{thebibliography}{200}
\bibitem{note} The non-magnetic phase at $T=0$ is believed to extend
also below $\alpha=0.5$, down to a critical value, $\alpha_c^\prime$,
corresponding to the transition to conventional N\'eel order
(for $S=1/2$,  $\alpha_c^\prime\simeq 0.4$).  
See also Refs.~\protect\cite{revgs,pig}.
\end{thebibliography}
\end{document}